\documentclass[sn-basic]{sn-jnl} 

\usepackage{graphicx}%
\usepackage{multirow}%
\usepackage{amsmath,amssymb,amsfonts}%
\usepackage{amsthm}%
\usepackage{mathrsfs}%
\usepackage[title]{appendix}%
\usepackage{xcolor}%
\usepackage{textcomp}%
\usepackage{manyfoot}%
\usepackage{booktabs}%
\usepackage{algorithm}%
\usepackage{algorithmicx}%
\usepackage{algpseudocode}%
\usepackage{listings}%

\graphicspath{{./}{figures/}}

\newcommand{\msun}{\rm M_{\rm \odot}}

\newcommand\mdotEdd{\dot{M}_{\rm Edd}}

\begin{document}

\title[What is the hard spectral state in X-ray binaries? Insights from GRRMHD accretion flows simulations and polarization of their X-ray
emission]{What is the hard spectral state in X-ray binaries? Insights from GRRMHD accretion flows simulations and polarization of their X-ray emission}


\author[1]{\fnm{M.} \sur{Moscibrodzka}}\email{m.moscibrodzka@astro.ru.nl}
\affil[1]{\orgdiv{Department of Astrophysics/IMAPP}, \orgname{Radboud
    University}, \orgaddress{P.O. Box 9010, 6500 GL Nijmegen, The Netherlands}}

\abstract{X-ray binaries are known to exhibit different spectral states
  which are often associated with different black hole accretion modes. The exact geometry and properties of these accretion modes is still uncertain. Recent IXPE measurements of linear polarization of X-ray emission in canonical X-ray binary system Cygnus
  X-1 allow us to test models for the hard spectral state of accretion in a unique
  way. We show that general relativistic radiative magnetohydrodynamic
  (GRRMHD) simulations of accreting stellar black hole in a hard X-ray state
  may be consistent with the new observational information. In the presented framework, where first-principle models
  have limited number of free parameters, the polarimetric X-ray observations
  put constraints on the viewing angle of the inner hot accretion flow.}


\maketitle

\section{Introduction}

The direct images of supermassive black holes in M87* and Sgr A* released by
the Event Horizon Telescope revealed ring structures which are
interpreted as advection dominated relativistically hot accretion flows in an immediate vicinity of the black hole event
horizons \citep{EHTC:2019,EHTC:2022}. These images allow us to test complex accretion flow models with
unprecedented precision. It is natural and timely to ask whether a black hole
accretion processes are universal and whether models constructed mostly for the supermassive
black holes apply to accreting stellar-mass holes found in our Galaxy.

Stellar-mass black holes accreting matter from a stellar companion (hereafter XRB) are known to display several distinct X-ray spectral states
characterized by different luminosities, X-ray spectral indices,
multiwavelength components and timing properties
\citep{fender:2012,belloni:2016}. The two main XRB spectral states are a
low-hard and a high-soft states.  In the low-hard state, the luminosity is
lower, the X-ray spectrum has a power-law shape and the source has a persistent
radio component. During the high-soft state, the luminosity rises closer the
Eddington limit, the spectrum is dominated by a thermal component and the
radio counterpart is quenched. Sources can be also found in an intermediate
state with thermal and power-law components both visible in the X-ray
spectrum. Since many years it is generally understood that the X-ray soft
component is produced by a thermal disk and the power-law component is emitted
by the so called ``corona'' which has some, not yet well understood, connection to a
relativistic jet emitting photons in the radio \citep{done:2007}. One of the
long-standing questions is what is the geometry and properties of this corona
and whether the corona is actually a compact jet or rather a part of the disk
and, if so, then how is the corona connected to the jet. All these questions are
difficult to answer solely observationally
because accretion flows in XRBs are spatially unresolved.

In this work we construct a more detailed model of the accretion flow in XRB in the low-hard
state using state-of-the-art general relativistic
radiation-magnetohydrodynamic (GRRMHD) simulations. 
Recent Imaging X-ray Polarimetry Explorer (IXPE) measurements of
polarization of X-ray emission in XRBs in a hard state
\citep[e.g.,][]{krawczynski:2022} opened up a unique opportunity to test accretion
flow scenarios in these systems as suggested already by \citet{lightman:1976}
and \citet{rees:1975}. In this work we show that the GRRMHD
simulations indicate consistency with these new observational constraints.

The paper is structured as follows. In Section~\ref{sec:model} we describe our
model for hard-state accretion. In Section~\ref{sec:sed} we show the
polarimetric properties of the model. We compare our modeling results to
observations and to other models in Section~\ref{sec:modelcomparison}. 
We discuss prospects and directions for future studies in Section~\ref{sec:future}.

\section{Accretion model for a hard spectral state}\label{sec:model}

We carry out GRRMHD simulations of low-hard state accretion flow onto
stellar mass black hole using first-principles GRRMHD code \texttt{ebhlight}
\citep{ryan:2017}. The state-of-the-art code couples evolution of plasma and
magnetic fields to the evolution of radiation field accounting for
radiative looses and radiative forces. The radiation model includes continuum
processes such as: synchrotron emission, bremsstrahlung emission,
self-absorptions and Compton scatterings of photons. Additionally, the code
follows ions and electron temperatures separately allowing for development of
two-temperature plasma flows \citep{ressler:2015}. The simulations include
evolution of radiation field and hence they are not scale free.  Inspired by
recently published results on X-ray polarization in XRB Cygnus X-1
\citep{krawczynski:2022}, we scale our model exactly to this system. In our
simulation we set black hole mass to $M_{\rm BH}=20\msun$ \citep{mj:2021} and
dimensionless spin of the black hole to $a_*=0.9375$
\citep{fabian:2012,gou:2014}. The GRRMHD simulations require specifying the
gas density and magnetic field strength via scaling parameter $\mathcal{M}_{\rm unit}$. We run two simulations with $\mathcal{M}_{\rm unit} = 1.5 \times 10^{11}$ and $ 2.5 \times 10^{11}$.
Two values of $\mathcal{M}_{\rm unit}$ are chosen to model emission for observers with low (disk face-on) and high (disk edge-on) viewing angles. The mass accretion rates in both models are at the level of $\dot{m}\equiv \dot{M}/\mdotEdd \sim {\rm few} \times 10^{-4}$. In X-rays both models are as bright as $L_{X} \approx 2 \times 10^{36}$ ergs/s consistent with luminosity of Cygnus X-1 during IXPE detection reported by \citet{krawczynski:2022}\footnote{The IXPE X-ray flux reported in \citealt{krawczynski:2022} is $F_X = 3.5-4.5 \times 10^{-9} {\rm ergs\, cm^{-2}\, s^{-1}}$. Given the distance to Cyg X-1, $d$=2.22 kpc, the source X-ray luminosity $L_X = 2-2.5\times10^{36} {\rm ergs \, s^{-1}}$. The reported photon index is $\Gamma \approx 1.62$ meaning that the X-ray part of the spectrum in $\nu L_\nu$ vs $\nu$ plots should have negative slope of $-0.38$.}. The Bolometric luminosities of our models are $L_{\rm Bol} \approx 10^{38} \,{\rm ergs\,s^{-1}}$ ($L_{\rm Bol}/L_{\rm Edd}\approx0.05$).

\begin{figure*}
  \centering
  \includegraphics[width=0.24\linewidth]{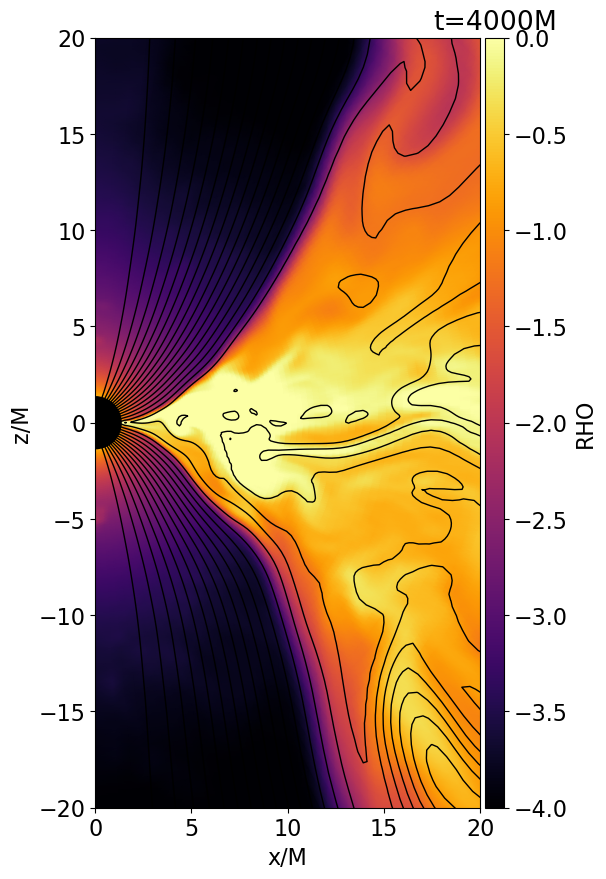}
  \includegraphics[width=0.24\linewidth]{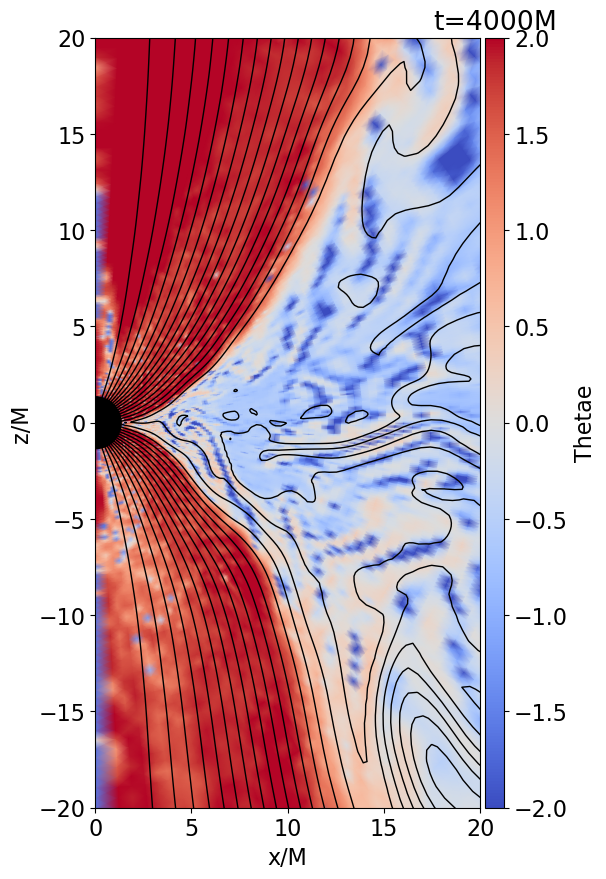}
  \includegraphics[width=0.24\linewidth]{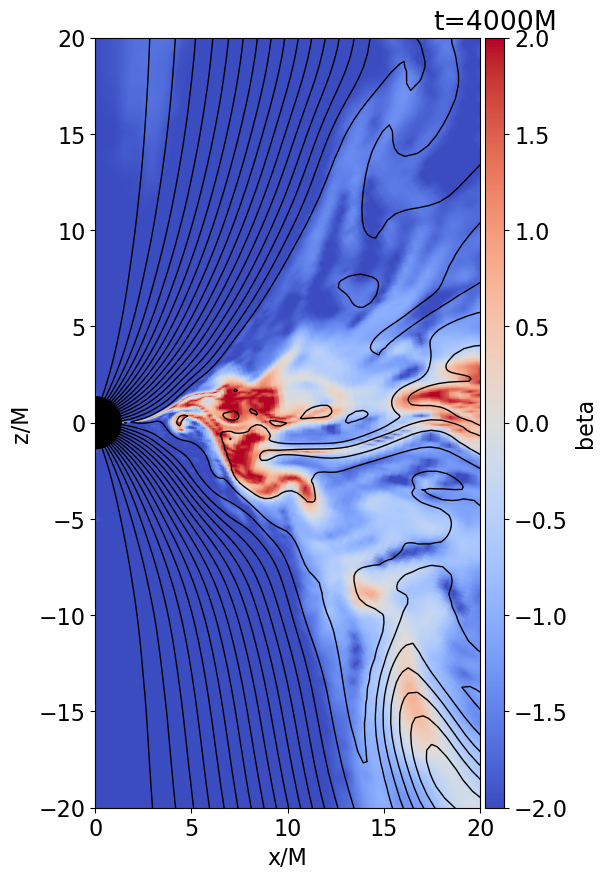}
  \includegraphics[width=0.24\linewidth]{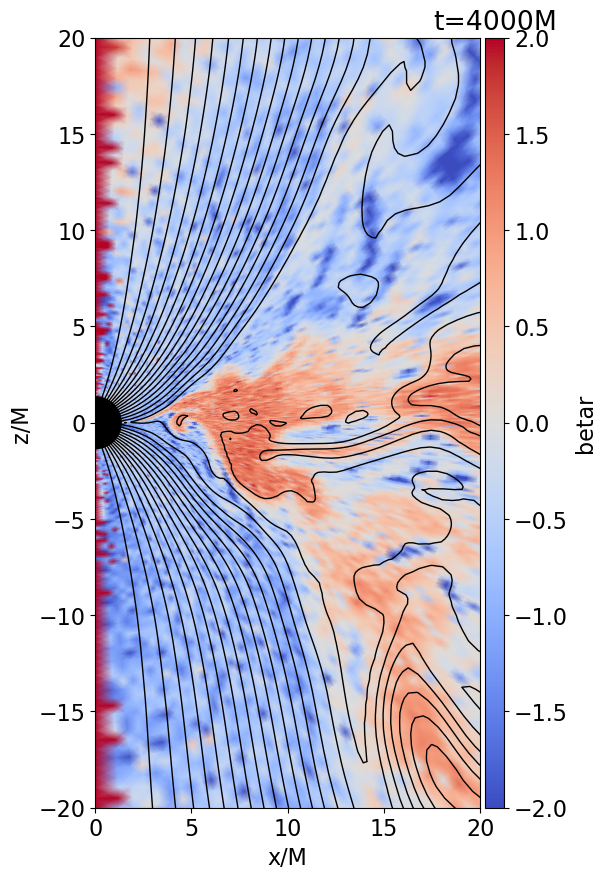}\\
  \caption{Geometry of the inner accretion flow in the hard spectral state in GRRMHD simulation with ${\mathcal M}_{\rm unit}=1.5 \times 10^{11}$ at time moment $t=4,000$M. Panels from left to right show the plasma density (in code units), the dimensionless electron
    temperature, the plasma $\beta$ parameter indicating dynamically important
    magnetic fields, and the radiation $\beta_r$ parameter (ratio of gas pressure to radiation
    pressure). The black contours follow the magnetic field lines.}\label{fig:remad}
\end{figure*}

The simulation initial conditions assume torus in hydrodynamical equilibrium
in the equatorial plane of the black hole. The torus is seeded with a loop of
poloidal magnetic field. Further details of the simulation setup are provided
in the Appendix~\ref{app:GRRMHD}. It is important to note here that our models initially do not contain a truncated geometrically thin, optically thick disk\footnote{Formally a thin disk could be self-consistently formed in the GRRMHD model inner region but our accretion rates are still too low for this to happen.} component \citep{ss:1973}. This means that in our models there is no reflection of the hot
accretion flow from the thermal thin disk and hence there is no thermal photons scattering off the hot corona. 
Our simulations self-consistently evolve magnetic fields. The magnetorotational instability leads to the
development of a turbulent accretion flow.
The accretion flow is accompanied by wind and a jet with organized magnetic fields.
The accretion flow models studied here are magnetically arrested (MADs,
\citealt{sasha:2011,mckinney:2012}, and see Appendix~\ref{app:GRRMHD}).
The powers of a jets produced by the MAD simulations are consistent with 
the estimated power of the relativistic jet observed in Cyg X-1 ($P_{\rm
  jet}=10^{36}-10^{37}$ergs/s, \citealt{gallo:2005,russell:2007}).

Fig.~\ref{fig:remad} presents maps of plasma density, electron temperature, plasma $\beta$
(gas to magnetic pressure) and radiation $\beta_r$ (gas to radiation pressure)
of the very inner accretion flow at a later
time moment of evolution. The very inner accretion flow is geometrically
thick,  it has two-temperature structure and it is rather strongly magnetised. 
The optical thickness of the simulations
is at most $\tau_{es} \approx n_e \kappa_{es} r_g \approx 0.03$ hence the radiation emitted from the gas may have impact on the gas via radiative back reaction rather than Compton forces. 
Interestingly, in the disk body the regions of colder electrons are spatially coincidental with low plasma $\beta$
regions suggesting that electrons are cooling rapidly in regions of stronger magnetic
fields. These sub-relativistic electrons may enhance Faraday rotation of polarized synchrotron photons (seed photons for Compton process) which may impact the polarization properties of the X-ray emission.

\section{Mock spectral energy distribution and polarization}\label{sec:sed}

Although the GRRMHD simulations do produce self-consistent emission spectra those spectra do not have polarization information. Hence, we post-process our GRRMHD
models with \texttt{radpol}, recently developed, fully relativistic, Monte Carlo code for
polarized radiative transfer \citep{moscibrodzka:2020,moscibrodzka:2022}. The
details of the calculation are provided in the Appendix~\ref{app:RT}.

Fig.~\ref{fig:radpol_sed} shows polarized broadband SED of the GRRMHD
models for a couple of viewing angles around times $t=4,000{\rm M}$. The lower viewing angle model has slightly higher ${\mathcal M}_{\rm unit}=2.5\times10^{11}$ to match the observed X-ray flux and spectral slope. At the shown time the simulation is already relaxed and approximately steady-state with variations due to magnetic turbulence.
Each SED has three components: synchrotron hump, X-ray scattered
emission and bremsstrahlung of which synchrotron hump and Compton humps are
the most prominent. The bottom panels in Fig.~\ref{fig:radpol_sed} show
polarimetric information. The fractional linear polarization
$|m|=\sqrt{Q^2+U^2}/I$ in X-ray band is small, less than 10\% for both viewing angles. The fractional linear polarization of the synchrotron emission is naturally higher. The electric vector position angle ($EVPA \equiv 0.5arg(Q+iU)$) is
zero for close to edge-on inclination angles, meaning that EVPA is aligned with the jet for
nearly edge-on view of the model. For lower viewing angles, the EVPA
significantly deviates from zero and displays some frequency dependency. 

Unfortunately, it is not straightforward to measure how much of the X-ray
polarization observed in the model is caused
by the scatterings or the seed photon polarization because one cannot disentangle
different effects easily in these complex simulations (e.g., the scattering process
does depend on the polarization of incident photons). Due to the complexity of the model it is even more difficult to explain the behaviour of the EVPA as a function of viewing angle and frequency.

\begin{figure*}
\centering
\includegraphics[width=0.98\linewidth]{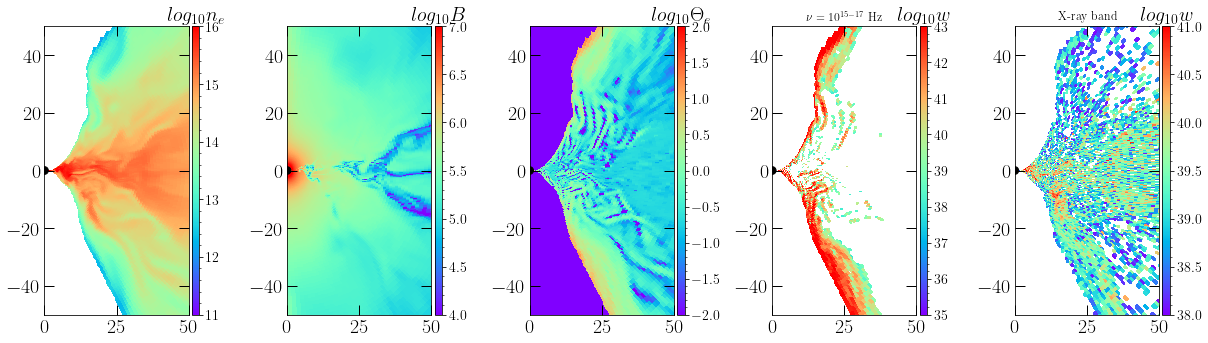}\\
\caption{Example maps of plasma density, magnetic field strength (both scaled to Cygnus X-1
  system and given in c.g.s. units), dimensionless electron temperature
  and maps of origin of photons at different energy bands in one snapshot of
  the GRRMHD simulation. The maps of the photon origin are $\phi$ angle
  averaged to improve signal-to-noise ratio in the maps. The color scale in
  the emission maps (two right-most panels) is proportional to the number of
  photons emitted. Notice that the size of the maps shown here is larger in comparison to Fig.~\ref{fig:remad}.}\label{fig:emiss_map}
\end{figure*}

In addition to the spectrum our radiative transfer simulations
track the origin of emission detected in different spectral windows.
In Fig.~\ref{fig:emiss_map} we show a snapshot of the GRRMHD simulation together with
emission maps produced by our \texttt{radpol} scheme. Notice that in comparison to
Fig.~\ref{fig:remad}, Fig.~\ref{fig:emiss_map} shows the model on 
larger scales (within $50$M) and plasma density and magnetic field strength
are shown in c.g.s. units. Here it is evident that the electrons in the GRRMHD
model are mildly relativistic with temperature increasing with
latitude. Hence the seed synchrotron emission is produced in the inner parts of the accretion disk but mostly along the jet funnel wall. The inverse-Compton scatterings
take place along the jet wall but also in the disk region.

\begin{figure*}
  \centering
  \includegraphics[width=0.45\linewidth]{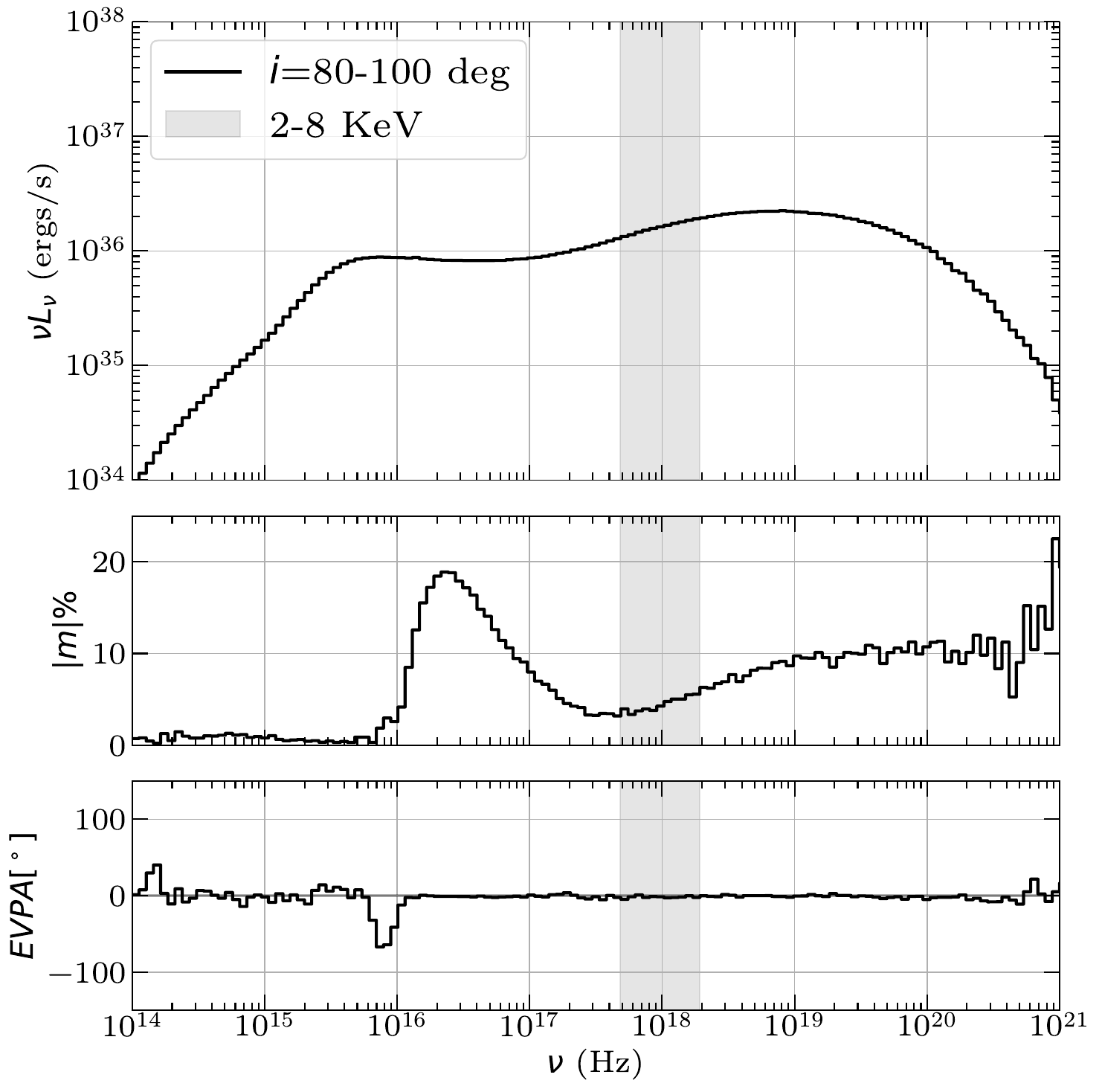}
  \includegraphics[width=0.45\linewidth]{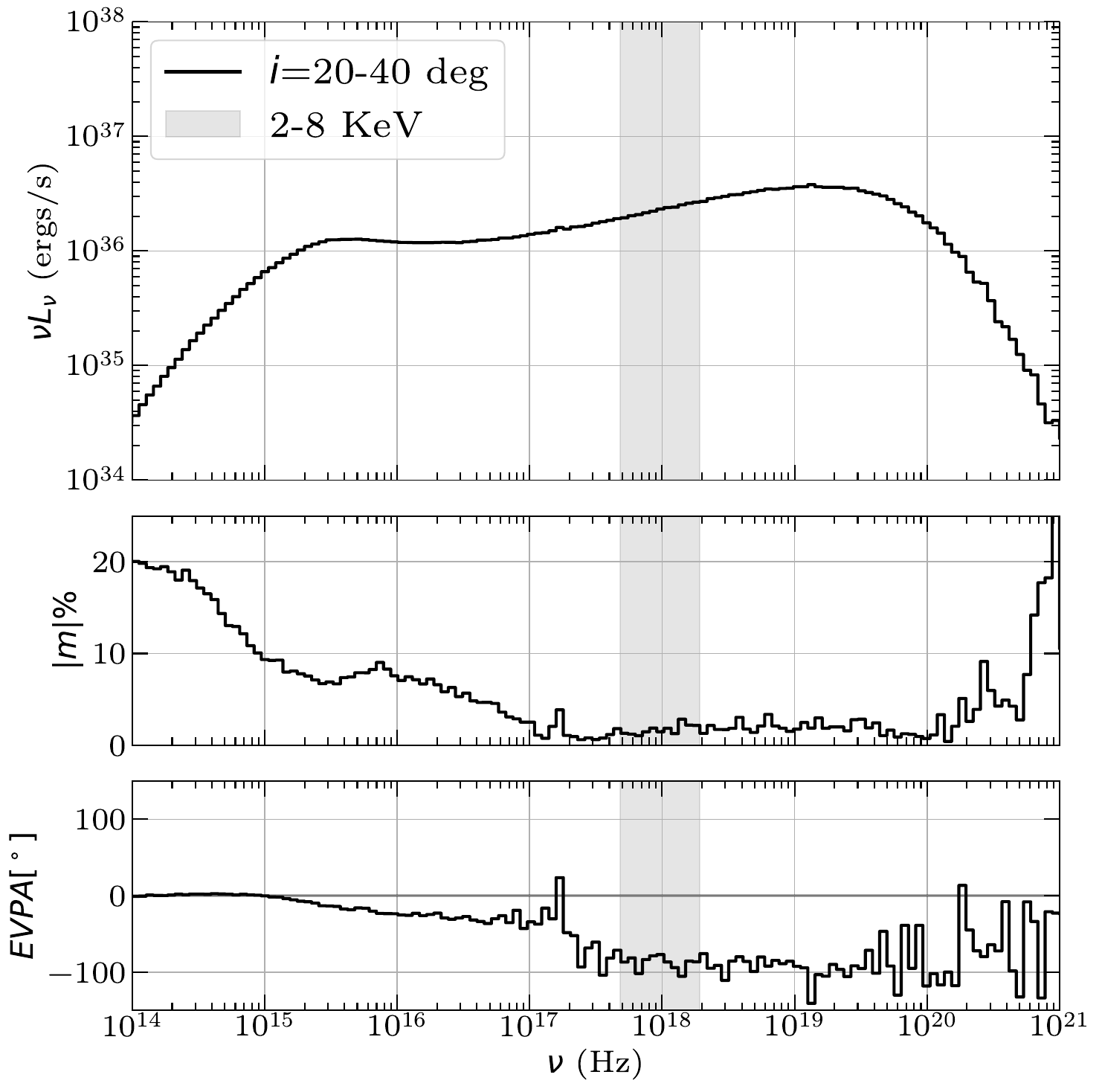}
  \caption{SEDs of the GRRMHD models at a viewing
    angle for which the SED matches observational constraints ($i=90$deg, left panel)
    and at a
    viewing angle equal to the estimated inclination angle of the binary
    system orbital plane ($i=30$deg, right panel). 
    Top panels show total
    intensity spectra, middle panels show the amplitude of the fractional
    linear polarization $|m|\%$ and lower panels display the linear
    polarization angle, EVPA. The gray shaded region marks bandwidth
    of the X-ray observations (2-8 keV).}\label{fig:radpol_sed}
\end{figure*}

\section{Comparison with observations and other models}\label{sec:modelcomparison}

The IXPE detected X-ray fractional polarization of Cygnus X-1 in hard state to be on
average $|m|=$4\% (slightly rising with frequency) and the EVPA to be aligned
with the radio jet position angle (in our reference system this translates to EVPA=0).
Given the simplifications in the model (e.g., the simulations are axisymmetric) our GRRMHD simulations of the
hard state are quite consistent with these observational constraints. Low polarization
fraction and EVPA aligned with the black hole spin or jet axis can be simultaneously recovered
by GRRMHD simulations assuming nearly edge-on viewing angles. In our model the X-ray emission
is emitted predominantly by the jet wall although some disk emission is also contributing.
For viewing angle of $i=30$deg (estimated viewing angle of the Cyg X-1 orbital plane, \citealt{mj:2021}) the EVPA can become nearly perpendicular (EVPA$\approx-100$ deg) compared to the observed
value. Overall our results indicate a possible misalignment between the jet axis (black hole spin axis) and the angular momentum of the binary or the outer accretion disk. 

In our model the X-ray emission and its polarization are produced by a partially
outflowing matter (jet wall). This is consistent with ideas recently put forward by others
\citep{poutanen:2023,dexter:2023} however the first-principles simulations
presented here prefer nearly edge-on viewing angles.  Our results are in
contradiction with the IXPE data interpretation presented in \citet{krawczynski:2022}
where authors argue that the measurement made in Cygnus X-1 can be interpreted
as a ``sandwich''-like corona above the thin disk or ``corona'' alone
(with either external Comptonization or self-synchrotron Comptonization)
observed at intermediate inclinations angles ($i=45-60$ deg). For such high viewing angles our model are depolarized in X-rays and EVPA deviates from zero. 
All these discrepancies can be attributed to different assumptions made when using different modeling techniques. For example, here we do not include the thin disk component, in contrast to the previous work. On the other hand, we include the previously
neglected Faraday rotation which very likely plays a role in depolarization
process of synchrotron emission in the disk where sub-relativistic electrons exist. Our
models also include self-consistently evolved magnetic field geometry which is
not included in the aforementioned geometrical models.
Alternatively, the current model may be incorrect and another GRMHD scenario should be considered to explain the IXPE measurements. 

\section{Future prospects}\label{sec:future}

Overall, the accurate detection of the X-ray polarimetry is a powerful new tool for constraining GRMHD simulations of XRBs and accreting supermassive black holes. In this work, we have made a first attempt to model polarimetric characteristics of XRB, utilizing two state-of-the-art numerical techniques. GRRMHD simulations, although already complex, still have several uncertainties that may impact the interpretation of the observations. At least two types of developments could be pursued. 

The sub-grid electron thermodynamics in the GRMHD simulations is uncertain. One could consider alternative to currently used electron heating model which assumes purely Alfv{\'e}nic cascade for energy dissipation.
However it is not clear which model for electron
heating is realistic for a given GRRMHD simulation 
\citep[e.g.,][]{howes:2010,kawazura:2018,satapathy:2023}. 
Another missing component in the GRRMHD models are the non-thermal electrons which are likely accelerated along the jet. Including the effect of acceleration using first-principles would require developing new modeling approaches.

A more feasible goal would be to calculate broadband polarization of a fully three dimensional simulations such as the ones presented in \citet{dexter:2021}. Our current model in an axisymmetric
version of fully three dimensional models M3/M4 in \citet{dexter:2021} but notice
that the our model differs in the black hole mass, evolution time and the size of the gas-radiation interaction zone (assumed to be 40M by \citet{dexter:2021}). Three
dimensional simulations allow to study emission from models with
accretion disk tilted with respect to the equatorial plane so they may be
helpful to explain the discrepancy between the viewing angle of the inner accretion flow and
the orbital plane of the system. Finally, only three dimensional GRRMHD models are suitable for time variability studies. 

\backmatter





\bmhead{Acknowledgments}
The author thanks Jiri Svoboda, Ma\l{}gosia Sobolewska, George Wong, Maciek Wielgus,
Pierre-Olivier Petrucci, Henric Krawczynski for useful discussions and
comments on observations of XRBs and simulations of accreting black
holes. Author also thanks Jason Dexter for discussion at early stage of the
project in 2021 and sharing one time slice of the 3-D model M4.
We gratefully acknowledge the HPC RIVR consortium (www.hpc-rivr.si) and EuroHPC JU (eurohpc-ju.europa.eu) for funding
this research by providing computing resources of the HPC system Vega at the Institute of Information Science (www.izum.si).










\begin{appendices}

\section{Methods}

\subsection{GRRMHD simulations}\label{app:GRRMHD}

The GRRMHD simulations of accretion onto a black hole are carried out by means of
public code
\texttt{ebhlight}\footnote{https://github.com/AFD-Illinois/ebhlight} developed
by \citet{ryan:2017}.  \texttt{ebhlight} integrates equations of ideal
radiation-magnetohydrodynamics in Kerr space time where multi-frequency
radiative transport is integrated out using Monte Carlo techniques. The code
additionally tracks total and electron entropy equations which allow to
calculate total and electron heating rates as the model evolves \citep{ressler:2015}. The latter
requires a model for distribution of dissipated energies between electrons and
protons. We adopt a sub-grid model for turbulent heating of electrons and ions (see e.g., \citealt{howes:2010}) as implemented in \texttt{ebhlight} by the code
authors. The electron model also accounts for Coulomb couplings and radiative
cooling.  The self-consistently evolved radiative transfer includes
synchrotron emission and absorption, Compton processes and bremsstrahlung. The \texttt{ebhlight} scheme allows for all possible matter-radiation interactions
(four-momentum exchange during emission, absorption and scatterings or, in
other words, cooling and radiative forces).

\begin{figure*}
  \centering
  \includegraphics[width=0.98\linewidth]{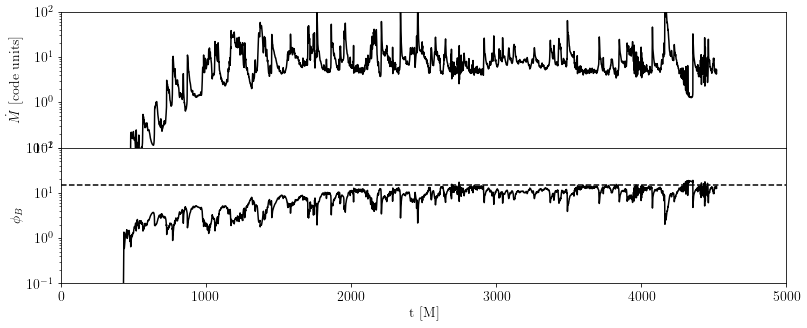}
  \caption{Evolution of the mass accretion rate ($\dot{M}$) and magnetic field
    flux ($\phi_{\rm B}\equiv \Phi_{\rm B}/\sqrt{\dot{M}c^2}$) near
    the event horizon of the black hole in our axisymmetric GRRMHD
    simulation.
    The magnetic field flux at the black hole horizon is normalized
  by the square root of the accretion rate at the horizon. The horizontal dashed line marks saturation level
  $\phi_B=15$ for which the accretion flows becomes magnetically arrested (MAD).}\label{fig:mdot}
\end{figure*}

Our initial conditions is the standard Fishbone and Moncrief torus, often
used to initialize advection dominated accretion flow models \citep{fishbone:1976}. The initial
torus is seeded with single loop of poloidal magnetic field.
The initial magnetic field configuration is defined using the vector potential:
\begin{equation}
(A_r,A_\theta,A_{\phi})=(0,0, \left(\frac{\rho}{\rho_{max}}\right) r^3 \sin^3(\theta) e^{-\frac{r}{400}} -0.2)
\end{equation}
where $r$ is the radius and $\theta$ is the polar angle in range from $0$ deg to
$180$ deg. This initial magnetic field configuration leads to the development of the Magnetically Arrested Disks (MADs, \citep{sasha:2011,mckinney:2012}). We integrate the MAD
solution using \texttt{ebhlight} until  $t_f=4,500$M, where
$M=GM/c^3$ is the time unit in geometriesed unit system ($G=c=1$).
Fig.~\ref{fig:mdot} shows evolution of mass accretion rate and magnetic flux
over time confirming large accumulation of the magnetic flux at the horizon
consistent with the MAD solutions obtained by other authors.

In the GRRMHD simulation, our grid is two dimensional and has resolution of $240\times 240$ zones in radial and
$\theta$ direction. We use modified Kerr Schild coordinates which are
logarithmic in radius, and mixed-modified in $\theta$ direction for a better
resolution of the inner part of the disk in the equatorial plane of the black
hole (for details see the \texttt{ebhlight} code paper).
We assume a ceiling, $\Theta_{e,max}=10^3$, and a floor, $\Theta_{e,min}=10^{-3}$, for electron temperatures through out the computational domain. The emissivity formulas for
synchrotron processes used in \texttt{ebhlight} are inaccurate for $\Theta_e<0.3$ therefore all
synchrotron emission from plasma with sub-relativistic electron temperature is
suppressed. However, the synchrotron emission from plasma with $\Theta_e<0.3$ is negligible. Compton scatterings, on the other hand, are allowed for all possible electron temperatures.
In the GRRMHD simulation, we allow for interactions between gas and radiation field only within
100M of the simulation domain. In the outer regions these interactions are negligible
when it comes to the gas dynamics. The outer radius of the simulation extends
to $R_{\rm out}=1000$M.

\subsection{Polarized radiative transfer simulations}\label{app:RT}

General relativistic polarized radiative transfer is carried out using
postprocessing Monte Carlo scheme \texttt{radpol} \citep{moscibrodzka:2020,moscibrodzka:2022}.
\texttt{radpol} includes polarized synchrotron emission, self-absorption and polarization
sensitive Compton scatterings.
In the postprocessing we make the following simplifications. We assume that
synchrotron emissivity for Stokes V and that Faraday conversion are null. These
processes are weak and not important for X-ray polarization predictions. (Notice
that the Faraday rotation is included and it affects the linear polarization.)
To be self-consistent with GRRMHD simulation, we assume that the electron distribution function is thermal.
Synchrotron emission from regions with $\Theta_e<0.3$ is suppressed similar to GRRMHD model,
but Compton scatterings on relativistically cold electrons are allowed, as well as Faraday
rotation. All postprocessing radiative transfer also makes a cut to
exclude uncertain emission from highly magnetized regions (defined with $\sigma \equiv B^2/ 4\pi \rho c^2 >1$) where \texttt{ebhlight} (similarly to all other GRMHD and GRRMHD codes) fails to accurately solve for internal gas energies). 
In the post-processing radiative transfer step we calculate emission including
entire GRRMHD computational domain also outside of 100M.
All presented spectra are $\phi$ angle averaged but are not folded around
equator.




\end{appendices}



\end{document}